# A Novel Framework to Expedite Systematic Reviews by Automatically Building Information Extraction Training Corpora

Tanmay Basu[1], Shraman Kumar[2], Abhishek Kalyan[2], Priyanka Jayaswal[2], Pawan Goyal[2], Stephen Pettifer[3] and Siddhartha R. Jonnalagadda[1]

[1] Division of Biomedical Informatics, Feinberg School of Medicine, Northwestern University, USA
[2] Department of Computer Science and Engineering, Indian Institute of Technology, Kharagpur, India
[3] School of Computer Science, University of Manchester, Manchester, UK

## Abstract

**Background:** A systematic review identifies and collates various clinical studies and compares data elements and results in order to provide an evidence based answer for a particular clinical question. The process is manual and involves lot of time. A tool to automate this process is lacking.

**Objective:** The aim of this work is to develop a framework using natural language processing and machine learning to build information extraction algorithms to identify data elements in a new primary publication, without having to go through the expensive task of manual annotation to build gold standards for each data element type.

**Method:** The system is developed in two stages. Initially, it uses information contained in existing systematic reviews to identify the sentences from the PDF files of the included references that contain specific data elements of interest using a modified Jaccard similarity measure. These sentences have been treated as labeled data. A Support Vector Machine (SVM) classifier is trained on this labeled data to extract data elements of interests from a new article.

**Results:** We conducted experiments on Cochrane Database systematic reviews related to congestive heart failure using inclusion criteria as an example data element. The empirical results show that the proposed system automatically identifies sentences containing the data element of interest with a high recall (93.75%) and reasonable precision (27.05% - which means the reviewers have to read only 3.7 sentences on average).

**Conclusions:** The empirical results suggest that the tool is retrieving valuable information from the reference articles, even when it is time-consuming to identify them manually. Thus we hope that the tool will be useful for automatic data extraction from biomedical research publications. The future scope of this work is to generalize this information framework for all types of systematic reviews

## Introduction

A systematic review identifies and collates various clinical studies, and compares data elements and results in order to provide an evidence-based answer for a particular clinical question [1]. Despite their widely acknowledged usefulness, the data extraction phase of the systematic review process is time-consuming [2]. Some reviews on the automation of systematic review process describe technologies needed for automating the overall process or individual steps. Tsafnat et al. [3] described four main tasks in systematic review: identifying the relevant studies, evaluating risk of bias in selected trials, synthesis of the evidence, and publishing the systematic reviews by generating human-readable text from trial reports. The use of text mining to reduce screening workload in systematic reviews has been discussed by Shemilt et al. [4]. Miwa et al. [5] proposed an active learning framework to reduce the workload in citation screening for inclusion in the systematic reviews. A classification system for screening articles for systematic review has been presented by Adeva et al. [6]. Choong et al. developed a system for automatic citation snowballing that recursively pursues relevant literature for helping in evidence retrieval for systematic reviews [7]. Kiritchenko et al. [8] developed ExaCT, a tool that assists users with locating and extracting key trial characteristics such as eligibility criteria, sample size, drug dosage, and primary outcomes from full-text journal articles. The major limitation of ExaCT is that it uses manually labelled training set for information extraction from a new article.

Unfortunately, none of these existing reviews and articles focus on automatic data extraction from a full text article [2]. In a recent study, Tsafnat et al. [1] have described the informatics systems that automate some of the tasks of systematic review and report systems for each stage of systematic review and while data extraction has been described as a task in their review, they only highlighted three studies as an acknowledgement of the ongoing work.

Moreover, as we have noted in our review on this topic [2], there is a dearth in gold standards for information extraction from literature for the purpose of systematic reviews. It has been discussed that only three out of 26 studies included in that systematic review, use a common corpus, namely 1000 medical abstracts from the PIBOSO corpus [2]. Unfortunately, even that corpus facilitates only classification of sentences into whether they contain one of the data elements corresponding to the PIBOSO categories. No two other studies shared the same gold standard or dataset for evaluation. State of the art text mining techniques use gold standards of documents with concepts annotated manually and then builds machine-learning systems. However, it will be expensive to build gold standards even for 50 standard data elements and practically impossible to build gold standards for hundreds of domain-specific data elements.

We, therefore, present a framework to leverage the wealth of human curated information already present in the existing systematic reviews. We propose to develop our training set by retrieving sentences from the full text articles that correspond to the manually extracted data elements present in the systematic reviews. Instead of having humans to go through the content of the full PDF article, we expedite this process by first developing a heuristics based system for automatically annotating included studies to assist the human annotator or directly use the outputs of this system for data elements where the accuracy is high enough. It is significantly lesser effort on the part of the humans.

The aim of this work is two-fold. The first objective is to develop a framework to generate gold standard using existing systematic reviews by applying natural language processing (NLP) techniques such as sentence similarity algorithms to automatically create training corpora from the references mentioned in the systematic reviews. The other objective is to use these gold standards to train machine learning classifiers for extracting data elements from a new study. A SVM classifier has been used towards this learning task.

## Methods

The proposed system is developed in two stages – training corpus (or gold standard) generation and data extraction using this training corpus. A graphical abstract of the proposed framework is presented in Figure 1.

**Gold Standard Generation:** Initially, different data elements (e.g., the inclusion criteria of a health care study) and their references are extracted from each systematic review. The synopsis of different data elements are generally mentioned in systematic reviews. Therefore the entire information about the data elements are extracted from the associated studies included in the systematic reviews. A modified Jaccard similarity measure is used to identify specific sentences from an included reference that represent individual data elements. The modified Jaccard similarity measure (Jac_Mod) finds the similarity between the sentence ($S_x$) that represents a data element (D) in a systematic review and each sentence ($S_y$) in an included reference as follows:

$$Jac\_Mod\ (S_x, S_y) = (number\ of\ common\ terms\ between\ S_x\ and\ S_y) / (number\ of\ terms\ in\ S_x)$$

Note that the modified Jaccard similarity measure ranges between [0, 1]. The sentences are sorted according to their similarity values and the top few sentences are considered as the positive instance for D. Note that a specific data element may be represented in the references by more than one sentence. If the difference of Jac_Mod values between the top-most sentence and the next few sentences lie between 0 and $\alpha$ (say), then all of these sentences are considered as positive instance for D. The sentences that have very low similarity values are considered as negative instance for D. A range of [0, $\beta$] of Jac_Mod scores has been considered for creating negative instances. Thus, the gold standard for each data element is developed by combining the positive and negative instances from all the included studies in two different groups. Here $\alpha$, $\beta$ are two different thresholds on Jac_Mod score and their value should lie in range [0, 1]. The values of $\alpha$ and $\beta$ are data dependent and could be fixed at runtime based on the characteristics of gold standard.

**Automatic Data Extraction:** In the second phase, a machine learning model has been developed, where this gold standard is used to train a classifier to determine whether a sentence contains the data element of interest (in our evaluation case – inclusion criteria) from a new clinical article. Unigrams, Bigrams and Trigrams generated from sentences are used as features with SVM classifier in the experiments. In its simplest form, SVM is a hyperplane that separates a set of positive examples from a set of negative examples with maximum margin [9]. Given a set of training documents in a vector space, SVM finds the best decision hyperplane that separates two categories. The quality of a decision hyperplane is determined by a margin between two hyperplanes that are parallel to the decision hyperplane and touch the closest documents of each category. Therefore best decision hyperplane is the one with the maximum margin, and is used to categorize the new test documents.

## Results

**Datasets:** 31 Cochrane Database systematic reviews (CDSR) related to congestive heart failure are randomly identified from PubMed. PDFs of the corresponding full-texts of the systematic reviews and the PDFs of the associated clinical studies are obtained automatically through APIs accessible from Northwestern University's Galter library for health sciences [10]. For obtaining the PDFs of the included references, the titles of these references are automatically extracted from the webpage of the corresponding systematic reviews in online Cochrane Library [11]. Therefore these titles are passed as individual queries to PubMed using eUtils webservice [12] to automatically retrieve the EndNotes of those references. These EndNotes are used to download the corresponding full-texts (in PDF format) through EndNote API of Galter library. The systematic reviews contain a section (typically called 'Characteristics of Studies' or a variation thereof) in which the properties of each clinical study are summarized by means of a table. A snapshot of a section of such table is shown in Figure 2 from the systematic review by Martí-Carvajal et al. [13]. In this figure, the keys, e.g., eligibility, inclusion criteria, exclusion criteria etc. along with their values are the data elements. Either the exact sentences or some significant information about these data elements are manually extracted from the referred articles and noted in such tables to develop systematic reviews.

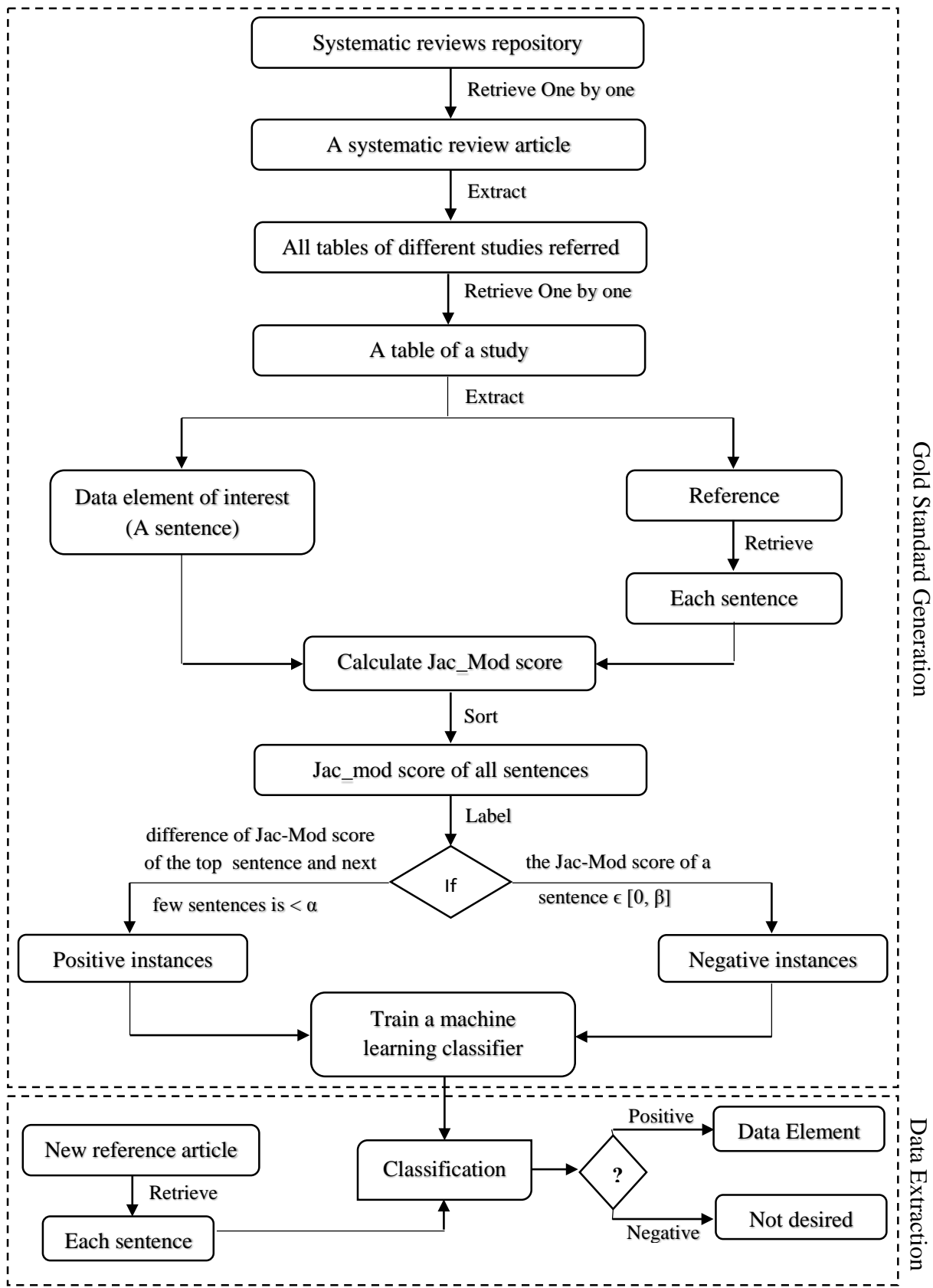

Figure 1: Graphical abstract of the proposed system

**Study Design:** The PDF files of each of these 31 systematic reviews are converted to JSON format using a version of the PDFx tool [14] customized for the structure of Cochrane Systematic Reviews. The JSON file format represents a machine readable version of these tables from which data elements can easily be retrieved programmatically. We have tested the performance of the proposed system on a commonly used data element (inclusion criteria of Randomized Control Trials (RCT)). Therefore the tables that contain inclusion criteria as data element are considered in the experiments. Eventually 69 such references have been considered to generate the gold standard. The PDF version of each of these references is converted to text format using a PDF to text converter tool freely available on internet.

The proposed system is implemented in scikit-learn, a machine learning platform developed in Python [15]. The gold standard is automatically developed following the proposed algorithm using these 69 references in scikit-learn. The values of α and β of Jac_Mod similarity measure have been fixed as 0.2 and 0.005 respectively. The inclusion criteria are extracted from each reference, which is considered as positive instance. After applying the method for gold standard generation described above for these 69 references, the gold standard contains 122 such positive instances and 12651 negative instances for inclusion criteria. Each instance represents a sentence. The test set contains 24 new RCT publications that are chosen randomly from the Cochrane Database [2] and are not included in the gold standard. These publications have been cited in some of the Cochrane Database Systematic Reviews. It has been confirmed that every reference contains some inclusion criteria based on them being explicitly mentioned in the corresponding systematic review. Therefore the inclusion criteria have been manually identified from these references.

| Methods | Multicentre study: yes<br>Country: USA<br>Intention-to-treat: yes (an intention-to-treat analysis was performed for all participants who had carotid ultrasonography at baseline and at least 1 follow-up visit, page 731)<br>Unit of randomization: patients<br>Follow-up period (years): B vitamins group (3.14 (0.48 to 4.56) versus placebo group (3.07 (0.46 to 5.0)) |
|---|---|

| Participants | Eligibility: 5309 |
|---|---|
| | Randomised: 506 (254 vitamins versus 252 placebo) |
| | Age (years) |
| | - Overall: 61.4 |
| | - B vitamins group: 61.7 ± 0.1) |
| | - Placebo group: 61.1 (±9.6) |
| | Gender (men): |
| | - Overall: 61% |
| | - B vitamins group: 61% |
| | - Placebo group: 61% |
| | Inclusion criteria: |
| | 1. Men and postmenopausal women 40 years old |
| | 2. Fasting tHcy 8.5 mol/L |
| | 3. No clinical signs/symptoms of cardiovascular disease (CVD) |
| | Exclusion criteria: |
| | 1. Fasting triglycerides > 5.64 mmol/L (500 mg/dL) |
| | 2. Diabetes mellitus or fasting serum glucose > 6.99 mmol/L (126 mg/dL) |
| | 3. Systolic blood pressure ≥ 160 mm Hg and/or diastolic blood pressure ≥ 100 mm Hg |
| | 4. Untreated thyroid disease |
| | 5. Creatinine clearance < 70 mL/min |
| | 6. Life-threatening illness with prognosis 5 years |
| | 7. 5 alcoholic drinks daily |
| Interventions | HLI-intervention: folic acid (5 mg), vitamin B12 (0.4 mg) and vitamin B6 (50 mg, daily supplementation) |
| | Control: placebo |
| | Treatment duration: initial 2.5-year treatment period was extended on average 1 to 2 years |

Figure 2: Snapshot of a part of systematic review table

The aim is to automatically identify the inclusion criteria in each of these new references using the proposed tool and evaluate the performance using the manually labeled information from systematic reviews. Each sentence of a new test reference is categorized to either positive (i.e., inclusion criteria) or negative instance by the system.

**Experimental Analysis:** The performance of the proposed tool has been evaluated by precision and recall measure [7]. The precision and recall can be calculated using the standard formulae as follow:

$$Precision = (True\ positives) / (True\ positives + False\ positives)$$
$$Recall = (True\ positives) / (True\ positives + False\ negatives)$$

Here, true positive stands for the number of sentences that are correctly retrieved as positive instances. False positive is the number of sentences that the tool marked as positive, but actually belong to the negative category. False negative is the number of sentences that are retrieved by the machine as negative instances, but actually belong to the positive category.

| Article Id | True Positive | False Negative | False Positive | Recall | Precision |
|---|---|---|---|---|---|
| Test Sample 1 | 1 | 0 | 1 | 1 | 0.50 |
| Test Sample 2 | 1 | 0 | 6 | 1 | 0.14 |
| Test Sample 3 | 1 | 0 | 2 | 1 | 0.33 |
| Test Sample 4 | 1 | 0 | 2 | 1 | 0.33 |
| Test Sample 5 | 1 | 0 | 3 | 1 | 0.25 |
| Test Sample 6 | 1 | 0 | 2 | 1 | 0.33 |
| Test Sample 7 | 1 | 0 | 7 | 1 | 0.12 |
| Test Sample 8 | 1 | 0 | 7 | 1 | 0.12 |
| Test Sample 9 | 2 | 0 | 6 | 1 | 0.25 |
| Test Sample 10 | 1 | 0 | 2 | 1 | 0.33 |
| Test Sample 11 | 1 | 0 | 5 | 1 | 0.17 |
| Test Sample 12 | 1 | 0 | 5 | 1 | 0.17 |
| Test Sample 13 | 1 | 0 | 7 | 1 | 0.12 |
| Test Sample 14 | 1 | 0 | 3 | 1 | 0.25 |
| Test Sample 15 | 1 | 0 | 3 | 1 | 0.25 |
| Test Sample 16 | 2 | 0 | 0 | 1 | 1 |
| Test Sample 17 | 2 | 2 | 4 | 0.5 | 0.33 |
| Test Sample 18 | 1 | 0 | 4 | 1 | 0.20 |
| Test Sample 19 | 1 | 0 | 5 | 1 | 0.17 |
| Test Sample 20 | 1 | 1 | 3 | 0.5 | 0.25 |
| Test Sample 21 | 1 | 1 | 4 | 0.5 | 0.20 |
| Test Sample 22 | 1 | 0 | 7 | 1 | 0.12 |
| Test Sample 23 | 1 | 0 | 2 | 1 | 0.33 |
| Test Sample 24 | 1 | 0 | 4 | 1 | 0.20 |

Table 1: Recall and precision score of all new test references

Note that the objective of the proposed tool is to achieve high accuracy in terms of identifying inclusion criteria (or other data elements in general) from new articles. Therefore, a high recall is desirable for the system. The recall and precision for each new test article is measured using the manually labeled inclusion criteria described above. The individual true positive, false positive, false negative, recall and precision values of each of the 24 new articles are presented in Table1.

It may be observed from Table 1 that the number of false negatives is zero for most of the test samples. It indicates that the proposed system is able to retrieve information of interest from new references with the framework being trained by automatically generated corpora. The aggregate recall and precision of all the test references are 0.93 and 0.27 respectively. A precision of 0.27 is still efficient because the reviewers now need to review only 1/0.27=3.7 sentences per article for the data elements as opposed to reading the entire article or training a corpus using completely new annotations. Thus the performance shows the effectiveness of the proposed tool.

## Discussion

**Principal Observations:** It has been suggested in earlier studies that SVM performs well for text classification [16, 17]. Hence, SVM is chosen as the classifier in the proposed system. Moreover, we have done extensive experiments using several other classifiers, such as Random Forests, Naïve Bayes using 10 fold cross validation on the training corpus and eventually SVM outperforms the other classifiers in terms of recall. The training corpus is imbalanced, since it contains many negative instances (12651) and few positive instances (122). Hence we have used balanced class weights to avoid the proportional influence of the negative class. It has been recommended in the study by Hsu et al. [17] that the linear kernel is useful, when the number of features is very large. It should be noted that the natural language texts generally produce large number of features [18]. Therefore linear kernel is used in the experiments.

In SVM, C is a regularization parameter that controls the trade-off between achieving a low error on the training data and minimizing the norm of the weights [19]. For large values of C, the optimization algorithm of SVM chooses a small margin separating hyperplane, if that hyperplane could correctly classify all the training instances. Conversely, if the value of C is too small then the optimization algorithm chooses a large margin separating hyperplane, resulting a high misclassification rate. It may be noted that the value of C lies in (0, ∞]. The goal is to identify a good C value, so that SVM can accurately predict unknown test samples i.e., to minimize the misclassification rate. Usually the value of C is determined by grid search on different class intervals of C values [17]. In grid search, SVM is implemented by performing k-fold cross validation on the training set using every C value in a given range. A finer grid search is performed on a narrow range on the C value that has achieved best accuracy. The process continues to refine C value up to third or fourth decimal place, which is predefined. For example, if the initial range is considered as [1, 10] then the subsequent finer ranges may be (4.1, 4.2… 4.9), (4.21, 422… 4.29), (4.261, 4.262… 4.269) etc. and a particular value from the last interval say, 4.263 could fixed as C. Eventually, this value of C is used for classifying the test samples. In k-fold cross-validation, the training set is randomly divided into k subsets of equal size. Sequentially one subset is tested using the classifier trained on the remaining *k*-1 subsets. Thus, each instance of the whole training set is predicted once and therefore the cross-validation accuracy is the percentage of data that are correctly classified. In our experiments, we have performed 10-fold cross validation on the training set to determine the value of C ranges in (0, 1000] and refined it up to fourth decimal place.

It has been observed from various publications included in the systematic reviews that the sentences which represent the inclusion criteria generally contain the keywords or phrases like, *eligibility, entry conditions, eligible patients,* etc. Sometimes the phrase *inclusion criteria* is explicitly mentioned before its description. Therefore these keywords/ phrases help to reduce the manual load. But, it is difficult to identify an inclusion criteria where it is not represented by such regular keywords. The following inclusion criterion has been extracted from test sample 7 [20] that is used in the experimental analysis.

- *This prospective cohort is composed by patients who are 18 years old or older, diagnosed as having chronic heart failure of ischemic or non-ischemic etiology, at least 3 months before inclusion into the registry.*

Although the sentence does not contain any discriminating keyword for inclusion criteria, the proposed system extracted this sentence from the reference. The most interesting aspect of the proposed framework, however, is that no knowledge about the meaning of "inclusion criteria" is used in generating the training corpus or building machine learning features.

The system is retrieving the data element of interest as well as some sentences similar to that data element from a new article, which results to a low precision score. The following sentences have been extracted by the proposed tool from test sample 18 [21] to identify inclusion criteria.

- *Inclusion criteria were diagnosed heart failure, either by echocardiography, radiographic evidence of pulmonary congestion or typical symptoms and signs of heart failure.*
- *Since the majority of patients hospitalized with heart failure were over 75 years of age, other limitations of recruitment were that patients were in an end-stage of heart failure or other severe disease or had cognitive dysfunction.*
- *A total of 161 patients met the inclusion criteria, 55 patients declined to participate mainly due to fatigue.*
- *According to age, sex and co-morbidities the patients included in our study are representative for the heart failure population.*

It should be noted that only the first sentence represents the inclusion criteria. The other sentences appear either in the results section or discussion section in test sample 18 [21], but they have some relevance with the first sentence. Hence the tool has extracted all of them as data elements, which results in a low precision score. The same scenario is observed in almost all the test samples. Note that the aggregate precision score of the system is reasonable (0.27), since the reviewers have to read only 3.7 sentences in a new article on average.

**Limitations:** The first limitation of the tool is that the study cannot guarantee that SVM will be useful in every case. It may be noted here that a particular classifier cannot provide an optimum solution in every situation and for any type of data [9]. Hence, the tool may show good results using other classifiers in machine learning for various other types of articles.

Moreover, the values of the thresholds α and β of Jac_Mod similarity measure should be chosen carefully to create the training corpus. It is recommended that both of the values should be close to 0. Otherwise many irrelevant sentences could be selected in the training corpus, which may degrade the performance of the machine learning classifier.

The performance of the tool is tested on some randomly selected articles from PubMed, mostly related to heart disease. The performance should be tested on studies from different topics of medical research e.g., internal medicine, oncology, neuroscience etc.

**Conclusions:** The literature survey has shown that a tool is lacking for automatic data extraction from a full text article. Therefore, a framework has been proposed in this study to overcome this limitation. It has been observed from the experiments that the tool is retrieving valuable information from some references, even when it is difficult to identify them manually. Thus, we hope that the tool will be useful for automatic data extraction from biomedical literature. The future scope of this work is to generalize this information extraction framework for all topics and to validate it for all frequently used data elements.

# References

1. Tsafnat G, Glasziou P, Choong MK, Dunn A, Galgani F, Coiera E. Systematic review automation technologies. Systematic Reviews; 2014; 3:74. DOI: 10.1186/2046-4053-3-74.
2. S. R. Jonnalagadda, P. Goyal and M. D. Huffman. Automating data extraction in systematic reviews: a systematic review. Systematic Reviews; 2015; 4:78. DOI: 10.1186/s13643-015-0066-7
3. Tsafnat G, Dunn A, Glasziou P, Coiera E. The automation of systematic reviews. BMJ; 2013; 346:f139. DOI: 10.1136/bmj.f139
4. Shemilt I, Simon A, Hollands GJ, Marteau TM, Ogilvie D, O'Mara-Eves A, et al. Pinpointing needles in giant haystacks: use of text mining to reduce impractical screening workload in extremely large scoping reviews. Research Synthesis Methods; 2014; 5(1):31-49. DOI: 10.1002/jrsm.1093.
5. Miwa MTJ, O'Mara-Eves A, Ananiadou S. Reducing systematic review workload through certainty-based screening. Journal of Biomedical Informatics; 2014; 51:242-253 DOI:org/10.1016/j.jbi.2014.06.005
6. Adeva JJG, Pikatza Atxa JM, Ubeda Carrillo M, Ansuategi ZE. Automatic text classification to support systematic reviews in medicine. Expert Systems and Applications; 2014; 41(4):1498-1508.
7. Choong MK, Galgani F, Dunn AG, Tsafnat G. Automatic Evidence Retrieval for Systematic Reviews. JMIR; 2014; 16:10. DOI: 10.2196/jmir.3369.
8. Kiritchenko S, de Bruijn B, Carini S, Martin J, Sim I. ExaCT: Automatic Extraction of Clinical Trial Characteristics from Journal Publications. BMC Med Inform Decis Mak. 2010;10:56.
9. Witten IH, frank E and Hall MA. Data Mining: Practical Machine Learning Tools and Techniques; 2011. ISBN-13: 978-0123748560.
10. Galter Health Sciences Library, Northwestern University, Chicago USA. https://galter.northwestern.edu/.
    Archived at: http://www.webcitation.org/6ghFMkbUL
11. Cochrane Library. http://www.cochranelibrary.com/.
    Archived at: http://www.webcitation.org/6gh0sjtMu
12. Sayers E, "A General Introduction to the E-utilities," 2009. http://www.ncbi.nlm.nih.gov/books/NBK25497/#_ncbi_dlg_citbx_NBK25497.
13. Martí-Carvajal AJ, Sola I, Lathyris D, Salanti G. Homocysteine lowering interventions for preventing cardiovascular events. Cochrane Database Systematic Review, 2015. DOI: 10.1002/14651858.CD006612.pub2.
14. Hager C. PDFx: Extract references (PDF, url, doi, arxiv) and metadata from a PDF. https://github.com/metachris/PDFx. Archived at: http://www.webcitation.org/6gFnv66VI

15. Pedregosa F, Varoquaux G et al. Scikit-learn: Machine Learning in Python, The Journal of Machine Learning Research; 2011; 12:2825-2830.
16. Joachims T. Text categorization with support vector machines: Learning with many relevant features. Proceedings of the European Conference on Machine Learning; 1998; p. 137-142.
17. Hsu CW, Chang CC, and Lin CJ. A Practical Guide to Support Vector Classification, 2010. http://www.csie.ntu.edu.tw/~cjlin/papers/guide/guide.pdf.
Archived at: http://www.webcitation.org/6gFnWKILv
18. Basu T, Murthy CA. A supervised term selection technique for effective text categorization. International Journal of Machine Learning and Cybernetics; 2015. DOI: 10.1007/s13042-015-0421-y
19. A user's guide to support vector machines, data mining techniques for the life sciences, Springer; 2009; p. 223-239. DOI: 10.1007/978-1-60327-241-4_13
20. Grancelli H, Varini S, Ferrante D, Schwartzman R, Zambrano C, Soifer S, Nul D, Doval H; GESICA Investigators. Randomized Trial of Telephone Intervention in Chronic Heart Failure (DIAL): study design and preliminary observations. Journal of Cardiac Failure; 2003; 9(3):172-179.
21. Stromberg A, Martensson J, Fridlund B, Levin LA, Karlsson JE, Dahlstrom U; Nurse-led heart failure clinics improve survival and self-care behaviour in patients with heart failure: results from a prospective, randomised trial. European Heart Journal; 2003; (11):1014-1023.